# Nonlinear quantum dot optomechanics

Matthias Weiß[1,2], Daniel Wigger[3,4], Maximilian Nägele[1], Kai Müller[5,2], Jonathan J. Finley[5,2], Tilmann Kuhn[4], Paweł Machnikowski[3], Hubert J. Krenner[1,2,*]

[1] Lehrstuhl für Experimentalphysik 1 and Augsburg Centre for Innovative Technologies (ACIT), Universität Augsburg, Universitätsstraße 1, 86159 Augsburg, Germany.

[2] Nanosystems Initiative Munich (NIM), Schellingstraße 4, 80799 München, Germany.

[3] Department of Theoretical Physics, Wrocław University of Science and Technology, 50-370 Wrocław, Poland.

[4] Institut für Festkörpertheorie, Universität Münster, Wilhelm-Klemm-Straße 10, 48149 Münster, Germany.

[5] Walter Schottky Institut, Technische Universität München, Am Coulombwall 4, 85748 Garching, Germany.

[*] corresponding author: hubert.krenner@physik.uni-augsburg.de

M. W. and D. W. contributed equally to this work.

**Abstract:**

Wave mixing is an archetypical phenomenon in bosonic systems. In optomechanics, the bi-directional conversion between electromagnetic waves or photons at optical frequencies and elastic waves or phonons at radio frequencies is building on precisely this fundamental principle. Surface acoustic waves provide a versatile interconnect on a chip and, thus, enable the optomechanical control of remote systems. Here, we report on the coherent nonlinear three-wave mixing between the coherent fields of two radio frequency surface acoustic waves and optical laser photons via the dipole transition of a single quantum dot exciton. In the resolved sideband regime, we demonstrate fundamental acoustic analogues of sum and difference frequency generation between the two SAWs and employ phase matching to deterministically enhance or suppress individual sidebands. This bi-directional transfer between the acoustic and optical domains is described by theory which fully takes into account direct and virtual multi-phonon processes. Finally, we show that the precision of the wave mixing is limited by the frequency accuracy of modern radio frequency electronics.

**Main text:**

The phenomenon of wave mixing is well-known in nonlinear optics[1] and has been widely employed in numerous other wave phenomena[2–5]. Also, in optomechanics, coherent transduction between optical and radio frequencies has been achieved[6–8], recently in the limit of single vibrational and optical quanta[8]. In this field, SAWs[9] provide a versatile bus enabling the control of remote systems on a chip, both in the classical[10,11] and the quantum domain[12]. Thus, SAWs are attractive to implement quantum gates and transduction in hybrid architectures[13–15] with most progress being made for superconducting artificial atoms[16–18] and spins of defect centers[19–21]. However, the relevant quantum mechanical transitions in these artificial atoms can be addressed by microwave photons or phonons that are of the same energy. For fully-fledged transduction to optical frequencies, quantum emitters in the relevant spectral domain are necessary. Here, semiconductor quantum dots (QDs) are attractive because their transitions are naturally in the visible and can be further tuned by size and composition[22], embedded in electrically active devices[23,24], and the emitted single photons[25] and entangled photon pairs[26] can be efficiently extracted by optical cavities[27,28] or interfaced with optical fibres[29]. Moreover, QDs couple to the dynamic strain of a SAW via the deformation potential[30–32]. Because SAW frequencies may even exceed the dot's natural linewidth, the emission spectrum splits up into a series of phononic sidebands (PSBs)[31]. These properties in combination with the ability to coherently address the optical transition with lasers[5,33] make QDs ideally suited for coherent SAW-based optomechanical schemes overcoming the presumed detrimental effect of phonons on the coherence of the exciton[34].

Here, we report on coherent nonlinear three-wave mixing between the coherent fields of two $\omega_{\text{SAW}}/2\pi \geq 1\,\text{GHz}$ SAWs and $\omega_{\text{opt}}/2\pi \approx 330\,\text{THz}$ optical laser photons via the dipole transition of a single QD exciton. To this end, the optical transition of this quantum emitter is dynamically modulated by up to two SAWs via deformation potential coupling and the exciton transition is simultaneously driven by a resonant laser. Because our system is operated in the resolved sideband regime[31,35], sum and difference frequency generation processes between the two SAWs occur which are directly observed in the scattered photon spectrum. This bi-directional transfer between acoustic and optical domains obeys phase matching conditions between the two SAWs enabling the deliberate enhancement or suppression of optical spectral components. We develop a full theoretical model considering both direct and virtual multi-phonon processes, which is found to be in excellent agreement with our experimental findings. Finally, we show that the precision of the wave mixing is limited by the frequency accuracy of modern radio frequency electronics.

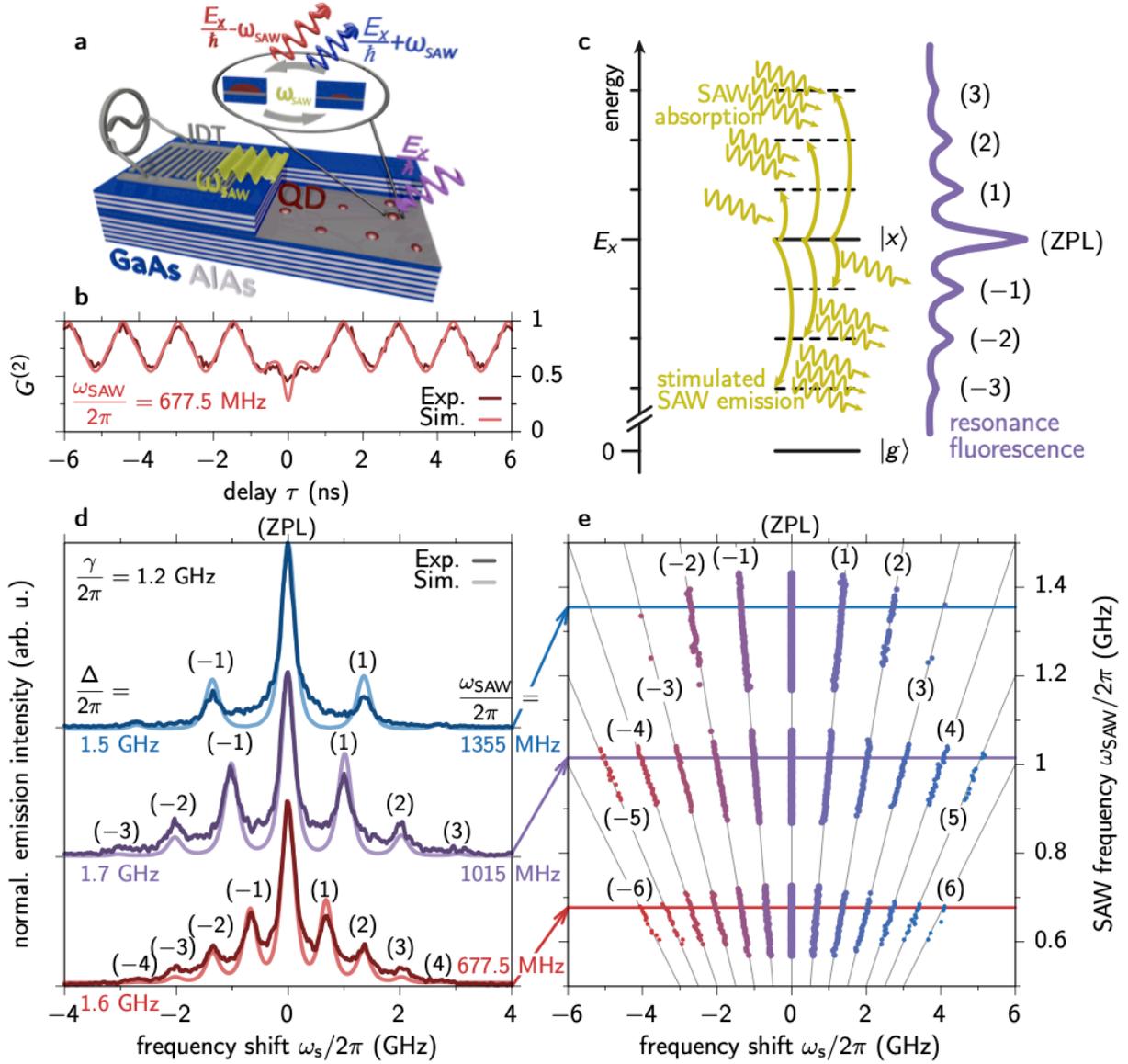

**Figure 1 | Frequency tunable PSBs. a,** Experimental implementation comprising a semiconductor QD embedded in a Bragg-type microcavity. The QD exciton transition is excited by a laser tuned close to or in resonance with its exciton transition at $E_X$. SAWs are generated by IDTs on the sample surface to dynamically strain the remotely positioned QD. **b,** Second-order correlation function $G^{(2)}(\tau)$ of a dynamically strained QD proving single photon statistics. **c,** Energy level diagram of the QD's SAW-modulated exciton transition being resonantly excited by a laser. Due to inelastic Stokes and anti-Stokes scattering processes, i.e. stimulated SAW phonon absorption and emission, the ZPL breaks up into a series of PSBs split by the phonon energy

$\hbar\omega_{\text{SAW}}$. **d,** Measured (dark lines) and calculated (light lines) scattered light spectra of a single QD dynamically strained by SAWs of three different SAW frequencies $\omega_{\text{SAW}}/2\pi$. The exciton decay $\gamma/2\pi$ and the SAW amplitudes $\Delta/2\pi$ used in the calculations are given in the plot. **e,** Measured PSB position as a function of electrical frequency, $\omega_{\text{rf}}$, applied to the IDT. The data shows nicely the three passbands of the IDT separated by narrow stop bands. The grey lines mark the expected $\omega_s = m \cdot \omega_{\text{SAW}}$ dependence of the PSBs.

The experimental setup for coherent optomechanical spectroscopy of single QDs is illustrated in Figure 1a. It comprises a GaAs/AlAs Bragg-type semiconductor microcavity with a layer of self-assembled In(Ga)As QDs at the antinode of the optical field. A multi-passband interdigital transducer (IDT), lithographically patterned on the sample surface, facilitates tunable excitation of SAWs of frequency, $\omega_{\text{SAW}}$, over four different frequency passbands. This is achieved by simply applying a radio frequency (rf) voltage, the frequency of which, $\omega_{\text{rf}}$, is faithfully mapped onto the frequency of the resulting SAW, $\omega_{\text{SAW}}$. Further details on the IDT designs and their characteristics are included in the Methods section and the Supplementary Information. In the following, only three of the four passbands are exploited in our experiment. The generated coherent acoustic field of the SAW propagates over distances $> 2000\ \mu\text{m}$ at the surface to the position of the optically interrogated QD. As shown in Figure 1a, the QD is dynamically strained by the SAW. The oscillating deformation potential induces an $\omega_{\text{SAW}}$-periodic spectral modulation of the optical transition[30,32,36] of amplitude $\hbar\Delta$ given by $E_X + \Delta E(t) = E_X + \hbar\Delta \cdot \cos(\omega_{\text{SAW}}t)$. For the SAW amplitudes applied in our experiment, any additional modulation due to a piezoelectrically induced Stark-shift[37] can be excluded. The exciton transition of a single QD is additionally excited by a narrow band continuous wave laser and the resonantly scattered photons are detected. Low optical pump powers are used to ensure that the coherence of the scattered single photons is determined by that of the laser[38]. First, we confirm the anti-bunched nature of the emitted light by measuring the second order correlation function $G^{(2)}(\tau)$ for the laser being detuned from the optical transition $E_X$. The experimental data plotted in dark red in Figure 1b shows two characteristic features, namely a $T_{\text{SAW}} = 2\pi/\omega_{\text{SAW}} = 1.476$ ns-periodic oscillation due to the dynamic modulation of the two-level system[35] and reduced coincidences, photon anti-bunching with $G^{(2)} < 0.5$ at $\tau = 0$ indicative for single photon emission. Other exemplary $G^{(2)}(\tau)$ measurements can be found in the Supplementary Information. Second, we study the system in the frequency domain where our system is set in the resolved sideband regime. To this end, we tune the driving laser in resonance with the QD's optical transition and employ this resonant excitation condition for all experiments presented in the remainder of the paper. With these settings, the narrow zero phonon line (ZPL) of the QD at energy $E_X$ splits into a series of phononic sidebands (PSBs) due to Stokes and anti-Stokes scattering. In the former and latter process, a discrete number of SAW phonons is emitted into or absorbed from the coherent bosonic field of the SAW, respectively. As shown schematically in Figure 1c, these PSBs are precisely spaced by the energy of the driving SAW phonon, $\hbar\omega_{SAW}$. Figure 1d shows typical scattered photon spectra of a single QD affected by SAWs with three different $\omega_{SAW}$. The recorded spectra are plotted as dark lines as a function of the frequency shift $\omega_s = \omega - E_X/\hbar$, relative to the QD's exciton transition. The spectra show the expected creation of a series of emission lines split precisely by the selected $\omega_{SAW}/2\pi$.[31,35] Our experimental data do not show any signatures of coherence loss such as line broadening. Moreover, the $G^{(2)}(\tau)$ experiment in Figure 1b and all spectra in Figure 1d, are well reproduced by simulations shown as light lines. In short, our theoretical model considers the QD exciton as an optically driven two-level system, described by the Hamiltonian $H = (E_X + \Delta E(t))\sigma_z + \frac{1}{2}\hbar\Omega(\sigma_+ + \sigma_-)$ with the time dependent shift of the transition energy $\Delta E(t) = \hbar\Delta \cdot \cos(\omega_{SAW}t)$. In these expressions, $\Omega$ denotes the Rabi frequency induced by the optical driving of the QD exciton and $\hbar\Delta$ the amplitude of the SAW modulation[30,36] introduced above. $\sigma_\pm = \sigma_x \pm i\sigma_y$ are the raising and lowering operators of the exciton two-level system with $\sigma_i$ denoting the Pauli spin matrices. Further details are included in the Methods section and the Supplementary Information. For the $G^{(2)}(\tau)$ data in Figure 1b we considered a laser detuning of $1.6 \cdot \hbar\omega_{SAW}$ and found the best agreement for a relative SAW amplitude of $D = \Delta/\omega_{SAW} = 3$. The quantities for Figure 1d are given next to each curve. Moreover, our theoretical model reproduces the experimental data almost perfectly using the same parameters as for the unmodulated QD. In particular, no additional pure dephasing rate is required. Owing to the multi-passband design of the employed transducers, the PSBs are fully tunable over a wide range of $\omega_{SAW}$ values. We demonstrate this in Figure 1e, where the measured positions of the PSBs are plotted as a function of the optical frequency shift $\omega_s$ and the SAW frequency $\omega_{SAW}$. Here, the IDT is driven by the electrical signal of frequency $\omega_{rf}$ and constant power $P_{rf} = +14\,\text{dBm}$ and generates SAWs with $\omega_{SAW} = \omega_{rf}$ across three of its wide frequency bands. These bands are separated by narrow stopbands where the IDT does not convert the electrical signal to a SAW. Details on the IDT design are provided in the Supplementary Material. Consequently, PSBs are observed at the programmed frequencies $\omega_s = m \cdot \omega_{SAW}$, with $m = 0, \pm 1, \pm 2, \ldots$ The number of resolved PSBs decreases with increasing $\omega_{rf}$ due to less efficient SAW generation and a reduction of the mean phonon number in the coherent SAW field, reflected formally by the reduction of the relative SAW amplitude $D = \Delta/\omega_{SAW}$. Clearly, the PSB positions faithfully reproduce $\omega_{rf} = \omega_{SAW}$ in the optical domain proving that the PSB spectrum can be easily tuned by changing $\omega_{rf}$.

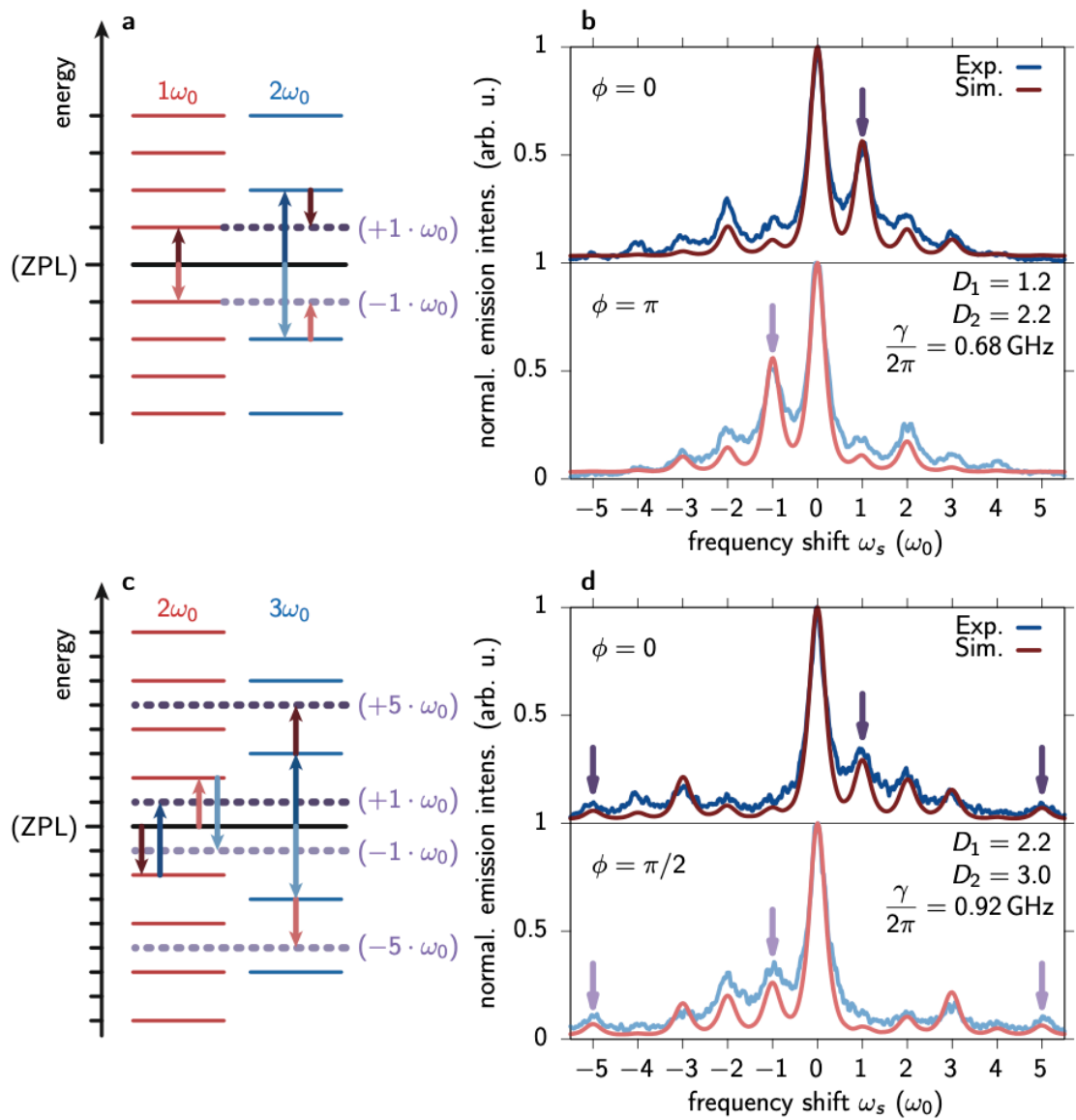

**Figure 2 | Sum and difference frequency generation. a,c,** Energy level diagrams indicating sum and difference SAW frequency generation processes for the

frequencies for $\omega_{SAW}^{(2)} = 2\omega_{SAW}^{(1)}$ in **a** and $2\omega_{SAW}^{(2)} = 3\omega_{SAW}^{(1)}$ in **c**. **b,d,** Measured (blue lines) and calculated (red lines) scattered light spectra corresponding to the two set SAW frequencies in **a,c**, respectively. By tuning the relative phase between the two SAW from $\phi = 0$ (upper panels) to $\phi = \pi$ (lower panel **b**) and $\phi = \pi/2$ (lower panel **d**) the intensities of different PSBs can be tuned due to phase matching.

In the next step, we demonstrate sum and difference frequency generation by using the QD exciton to mix two SAWs and the optical field, thus leading to an optically induced acoustic nonlinearity in an otherwise linear medium. To this end, we generate two SAWs with frequencies $\omega_{SAW}^{(1)}$ and $\omega_{SAW}^{(2)}$ as described in the Methods section and record the scattered photon spectrum. In Figure 2, we present two distinct cases. In the first case, the frequency of the second SAW is precisely twice that of the first SAW, i.e. $\omega_{SAW}^{(1)} = 1\omega_0$ and $\omega_{SAW}^{(2)} = 2\omega_0$, where $\omega_0$ is the reference frequency. In the second case, we chose $\omega_{SAW}^{(1)} = 2\omega_0$ and $\omega_{SAW}^{(2)} = 3\omega_0$. We begin with $\omega_{SAW}^{(1)}/2\pi = 0.6775 \text{ GHz}$ and $\omega_{SAW}^{(2)}/2\pi = 1.3550 \text{ GHz} = 2\omega_0/2\pi$ in Figures 2a,b, a scenario which is an acoustic equivalent to second harmonic generation in nonlinear optics. As shown in the level diagram in Figure 2a, we expect that the $+1$ and $-1 \cdot \omega_0$ PSBs can be generated by a set of processes combining well defined numbers of $1\omega_0$ (red arrows) and $2\omega_0$ (blue arrows) phonons. Up and down arrows indicate phonon absorption (anti-Stokes) and phonon emission (Stokes) processes, respectively. Combinations resulting in optical emission into the $+1$ or $-1 \cdot \omega_0$ PSB are indicated by dark and light arrows, respectively. Obviously, there exists the established single phonon processes as for the case of a single SAW field shown in the left part. The next higher processes generating the $\pm 1 \cdot \omega_0$ PSBs requires two phonons. Specifically, the $+1 \cdot \omega_0$ PSB is generated by the process marked by dark arrows in Figure 2a. It comprises absorption of a single $\omega_{SAW}^{(2)} = 2\omega_0$ phonon and emission of a single $\omega_{SAW}^{(1)} = 1\omega_0$ phonon. The underlying sum and difference frequency generation has to obey a phase matching condition because it occurs due to coherent wave mixing. This means that the intensity in each sideband depends on the relative phase $\phi$ of the two SAWs. In the following we reference this relative phase such that for $\phi = 0$ the two-phonon process described above is constructive. The opposite process, comprising the emission of one $\omega_{SAW}^{(2)}$ phonon and the absorption of one $\omega_{SAW}^{(1)}$ phonon, is marked by the light arrows in the figure and leads to pronounced emission into the $-1 \cdot \omega_0$ PSB. When $\phi = \pi$, the latter process is enhanced while the process dominating for $\phi = 0$ leading to $+1 \cdot \omega_0$ is suppressed. This expected phase matching is confirmed in our experiment, in which both frequency components are simultaneously generated by applying the two driving electric signals with the corresponding frequencies of powers $P_{\rm rf}^{(1)} = -6\,{\rm dBm}$ and $P_{\rm rf}^{(2)} = -2\,{\rm dBm}$ to the same IDT. In the data plotted as blue lines in Figure 2b, we clearly observe the expected anticorrelated enhancement and suppression of the two phonon frequency mixing processes leading to emission into the $+1$ or $-1 \cdot \omega_0$ PSB. For $\phi = 0$ (upper panel) the $+1 \cdot \omega_0$ PSB marked by the arrow is the dominant PSB while the $-1 \cdot \omega_0$ PSB is almost completely suppressed. When we set $\phi = \pi$ (lower panel), the situation is completely reversed and the $-1 \cdot \omega_0$ PSB (marked by arrow) is enhanced while the $+1 \cdot \omega_0$ PSB vanishes. We note that the asymmetry of the spectra is a direct evidence for nonlinear and coherent mixing in our case of the two bosonic acoustic fields and the laser's optical field through the QD's optical dipole. Furthermore, it is an acoustic analogue to the well-known phase matching in nonlinear optics and to the elimination of spectral components by all-optical quantum interference of two laser fields in an optical two-level system[39,40]. Note in case of incoherent mixing symmetric spectra are expected with the combined PSBs' intensities being the sum of that when the SAWs are applied individually. The above sum and difference frequency generation is further corroborated by data for two SAW fields with commensurate frequencies $\omega_{\rm SAW}^{(1)}$ and $\omega_{\rm SAW}^{(2)}$ such that $\omega_{\rm SAW}^{(1)}/2\pi = 2\omega_0/2\pi = 0.92\,{\rm GHz}$ and $\omega_{\rm SAW}^{(2)}/2\pi = 3\omega_0/2\pi = 1.38\,{\rm GHz}$. For this more general setting, no single frequency SAW is generated at $\omega_{\rm SAW} = \omega_0$ and, thus, emission of the $\omega_s = \pm 1 \cdot \omega_0$ PSBs is only possible via two-phonon processes with $\omega_s = \pm(\omega_{\rm SAW}^{(2)} - \omega_{\rm SAW}^{(1)})$. Similar holds for the $\pm 5 \cdot \omega_0$ PSBs for which $\omega_s = \pm(\omega_{\rm SAW}^{(2)} + \omega_{\rm SAW}^{(1)})$ has to be fulfilled. In the level diagram in Figure 2c, these processes are marked by arrows and the single frequency forbidden PSBs are indicated by the dashed lines. Again, in experiment both frequency components are generated by electrical signals ($P_{\rm rf}^{(1)} = 2\,{\rm dBm}$ and $P_{\rm rf}^{(2)} = 14\,{\rm dBm}$) applied to the same IDT. In the measured scattered photon spectra plotted in blue in Figure 2d, emissions of the $\pm 1 \cdot \omega_0$ and $\pm 5 \cdot \omega_0$ PSBs are clearly resolved. This observation is unambiguous and direct evidence of the coherent nonlinear sum and difference frequency generation of the two SAWs. Furthermore, the relative intensities of all PSBs is again fully tunable via the relative phase $\phi$ between the two SAWs. The upper and lower panels compare experimental spectra plotted in red for $\phi = 0$ and $\phi = \pi/2$, respectively. Again, similar to the data in Figure 2b, the curves are asymmetric manifesting coherent wave mixing. Interestingly, for the selected phases a pronounced asymmetry exists for the $\pm 1 \cdot \omega_0$ PSBs while in contrast, the $\pm 5 \cdot \omega_0$ PSBs exhibit similar intensities as indicated by the arrows. We extended our theoretical model to treat the experimentally employed dual acoustic modulation using relative SAW amplitudes $D_1 = \Delta^{(1)}/\omega_{SAW}^{(1)}$ and $D_2 = \Delta^{(2)}/\omega_{SAW}^{(2)}$, and a spontaneous emission rate $\gamma/2\pi$ as given in the plot. The relative phase $\phi$ is defined into the SAW modulation with the larger frequency. The simulated spectra are plotted in red in Figures 2b,d and faithfully reproduce the experimental data. Most strikingly, the phase matching underlying the different sum and difference frequency generation processes are in excellent agreement.

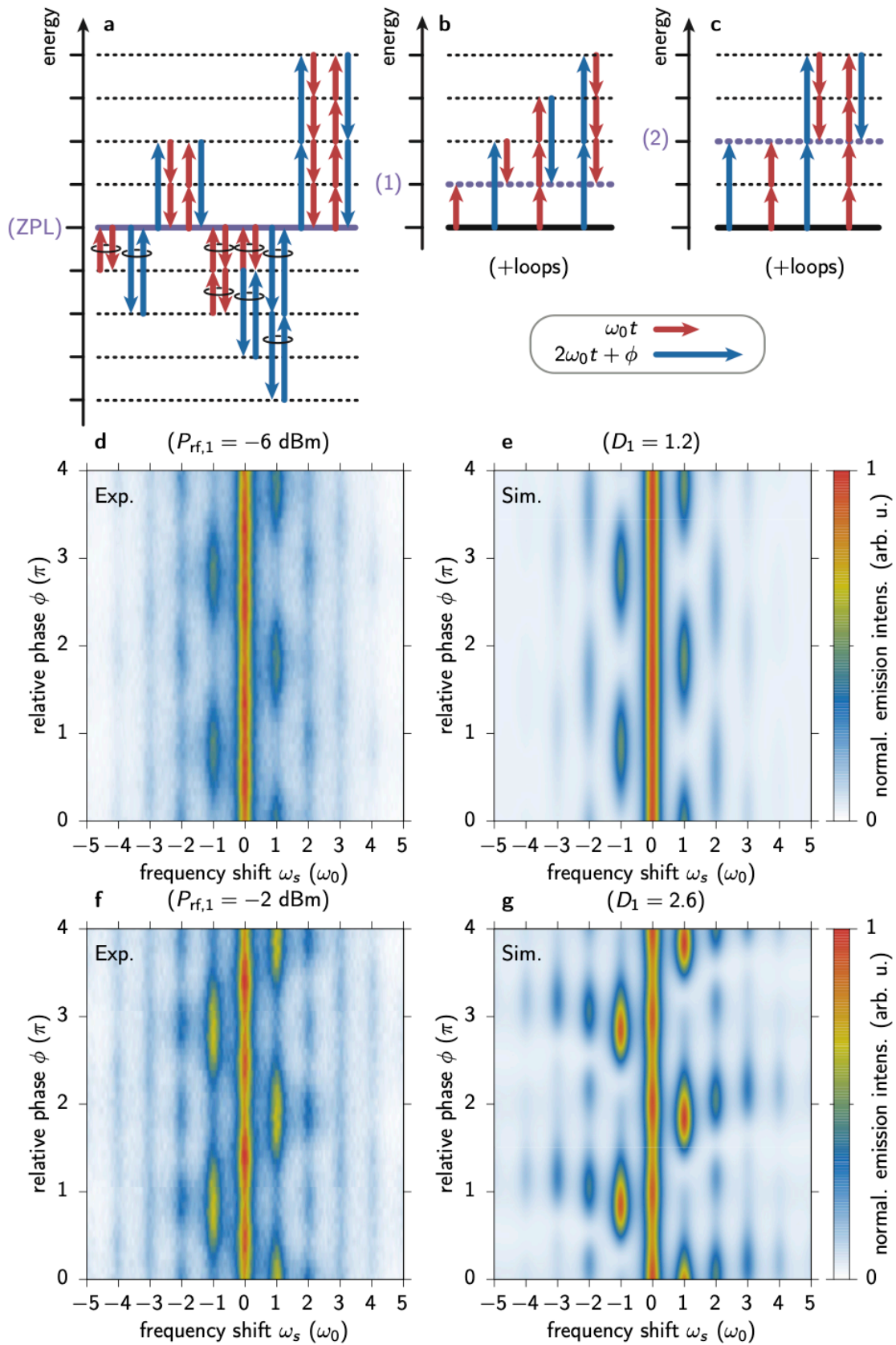

**Figure 3 | Phasematching. a,b,c,** Energy level diagrams indicating the lowest $\omega_{SAW}^{(1)} = \omega_0$ (red) and $\omega_{SAW}^{(2)} = 2\omega_0$ (blue) phonon absorption (up arrows) and emission (down arrows) processes leading to photon emission into the ZPL (a), the first (b) and second

PSBs (c), respectively. Virtual loop processes are indicated by coupled up and down arrows. All processes including loops are omitted in b and c. **d,e,** Relative phase dependence of the PSB spectrum in experiment and simulation, respectively. The normalized detected and calculated light intensities are color coded and plotted as a function of $\omega_s$ and $\phi$. **f,g,** Same as **d,e** but for a larger amplitude of the $1 \cdot \omega_0$ SAW as given in the plot.

In the next step, we address the phase matching ($\phi$-dependence) and higher order wave mixing processes for the case of $\omega_{SAW}^{(2)} = 2\omega_{SAW}^{(1)}$ both in experiment and theory. An analogous, supporting analysis for the $2\omega_{SAW}^{(2)} = 3\omega_{SAW}^{(1)}$ case is included in the Supplementary Information. We begin with higher order processes giving rise to emission into the ZPL, the $+1 \cdot \omega_0$ and the $+2 \cdot \omega_0$ PSBs. Figures 3a-c show the corresponding level diagrams where the $\omega_{SAW}^{(1)}$ transitions are marked as red arrows and those with $\omega_{SAW}^{(2)}$ phonons as blue arrows. From left to right the number of involved phonons increases. Importantly, an $\omega_{SAW}^{(2)}$ process carries a relative phase $\phi$ with respect to an $\omega_{SAW}^{(1)}$ process, as indicated in the Figure. Figure 3a shows the transitions giving rise to emission at the ZPL. Since this signal requires that the energies of the involved phonons precisely cancel, at least two phonons either $\omega_{SAW}^{(1)}$ or $\omega_{SAW}^{(2)}$ are required. In the coherent mixing process shown on the left for both frequencies, two phonons form a loop with one phonon being emitted and one phonon being absorbed. In these loops, phonon emission and absorption contribute relative phases of opposite sign. Thus, this precise cancellation ensures that such loops only contribute to the phase dependence by a constant amplitude. This is no longer the case for the next higher processes which involve three phonons. As shown in the second process of Figure 3a one $\omega_{SAW}^{(2)}$ phonon is absorbed (blue up arrow) and two $\omega_{SAW}^{(1)}$ phonons are emitted (red down arrows) or vice versa. These processes do not involve any loops but a single $\omega_{SAW}^{(2)}$ process and, therefore, contribute a phase $\phi$. The next higher processes are also shown in the level diagram involve two loops, i.e. four phonons. Based on these arguments, we do not expect a pronounced $\phi$-modulation of the ZPL for low SAW amplitudes since such a modulation requires at least three phonons. In Figure 3b we get back to the modulation of the $+1 \cdot \omega_0$ PSB and consider higher order processes contributing to emission into this PSB. The next higher order transition after that discussed in context of Figure 2 (absorption of $\omega_{\text{SAW}}^{(2)}$ and emission of $\omega_{\text{SAW}}^{(1)}$) involves four phonons, emission of a single $\omega_{\text{SAW}}^{(2)}$ and absorption of three $\omega_{\text{SAW}}^{(1)}$ quanta. This process is followed by that shown in the right of Figure 3b formed by absorption of two $\omega_{\text{SAW}}^{(2)}$ and emission of three $\omega_{\text{SAW}}^{(1)}$ quanta. Finally, Figure 3c shows an analogous consideration for emission into the $+2 \cdot \omega_0$ PSB. The lowest possible process is the absorption of an $\omega_{\text{SAW}}^{(2)}$ phonon. The next higher multi-phonon processes resulting in a modulation of this PSB require four or five phonons, namely the process of the absorption of two $\omega_{\text{SAW}}^{(2)}$ and emission of two $\omega_{\text{SAW}}^{(1)}$ quanta and that of emission of one $\omega_{\text{SAW}}^{(2)}$ and absorption of four $\omega_{\text{SAW}}^{(1)}$ quanta, respectively. These considerations show that the phase dependences of the different PSBs change when increasing the SAW amplitudes as higher phonon processes become more likely. Note, that there exist several higher processes for emission of the $+1 \cdot \omega_0$ and $+2 \cdot \omega_0$ PSBs which include loops introduced in the discussion of the ZPL-modulation. These loops are omitted in Figures 3b,c since they do not contribute a net phase. However, they are fully considered in our theoretical model and included in a Figure with the complete set of transitions in the Supplementary Material.

Next we study the discussed higher order processes in experiment and theory. We begin with the case of weak acoustic modulation. In Figure 3d, we present the measured optical emission of a single QD dynamically strained by two SAWs excited with powers of $P_{\text{rf}}^{(1)} = -6\,\text{dBm}$ and $P_{\text{rf}}^{(2)} = -2\,\text{dBm}$. In this regime, we expect that processes involving a single phonon of each frequency dominate. The normalized emission intensity is colour coded and plotted as a function of $\omega_s$ (horizontal axis) and $\phi$ tuned over two full periods, $[0,4\pi[$ (vertical axis). As expected, the $+1$ and $-1 \cdot \omega_0$ PSBs are the dominant PSBs. The intensities of these two PSBs exhibit precisely the expected $2\pi$-periodic modulation. They are also perfectly anti-correlated when $\phi$ is tuned, which follows from the interplay between $\omega_{\text{SAW}}^{(1)}$ and $\omega_{\text{SAW}}^{(2)}$ components, as explained in the Supplementary Information. Moreover, the ZPL's intensity remains almost constant when $\phi$ is scanned. As detailed above, this is expected because at least three phonons are required to induce a $\phi$-dependence. The $+2$ and $-2 \cdot \omega_0$ PSBs are less pronounced and are slightly modulated with a $2\pi$-periodicity stemming from the trivial single $\omega_{\text{SAW}}^{(2)}$ phonon process. Next, we compare the experimental data to our theoretical model which includes all possible multi-phonon processes contributing to the wave mixing. These processes include in particular those shown in Figure 3a-c. A complete overview of the relevant processes up to fifth order is included in the Supplementary Material. Figure 3e shows the results of the simulations using normalized SAW amplitudes $D^{(1)} = \Delta^{(1)}/\omega_{SAW}^{(1)} = 1.2$ and $D^{(2)} = \Delta^{(2)}/\omega_{SAW}^{(2)} = 1.5$ and an exciton decay rate $\gamma/2\pi = 0.81$ GHz. To better compare the results of our simulation to the experimental data, the simulation data in Figure 3e is plotted in the same way as the experimental data in Figure 3d. Clearly, the result of our simulation is in excellent agreement with the experimental data and fully reproduces all characteristic features. In particular the phase matching condition for emission into the $+1$ or $-1 \cdot \omega_0$ PSB is faithfully reproduced by our model.

The case of higher SAW amplitudes is studied in Figures 3f (experiment) and 3g (theory). In experiment, we increased $P_{rf}^{(1)}$ to $-2$ dBm and kept $P_{rf}^{(2)} = -2$ dBm unchanged compared to the first case. Under these experimental conditions, the ZPL now exhibits a clear $\pi$-periodic modulation. This type of modulation is in fact expected for the combination of the two 3-phonon processes in Figure 3a, in which each $\omega_{SAW}^{(2)}$ phonons contributes a $\phi$-dependence. In the same way the additional structure on the $\pm 2 \cdot \omega_0$ PSB can be understood. In the next order, a three-phonon process with one $\omega_{SAW}^{(2)}$ phonon and an $\omega_{SAW}^{(1)}$ loop also leads to the $2\pi$-periodicity also weakly resolved in Figures 3d and e. However, for larger SAW amplitudes, the four-phonon process, absorption (emission) of two $\omega_{SAW}^{(2)}$ phonons and emission (absorption) of two $\omega_{SAW}^{(1)}$ phonons, contributes significantly. This process involves two $\omega_{SAW}^{(2)}$ phonons which gives rise to the observed $\pi$-periodicity. Again, the corresponding simulations using our theoretical model ($D^{(1)} = \Delta^{(1)}/\omega_{SAW}^{(1)} = 2.6$, $D^{(2)} = \Delta^{(2)}/\omega_{SAW}^{(2)} = 1.5$) shown in Figure 3g nicely reproduce all characteristic $\pi$- and $2\pi$-periodic features of the experimental data.

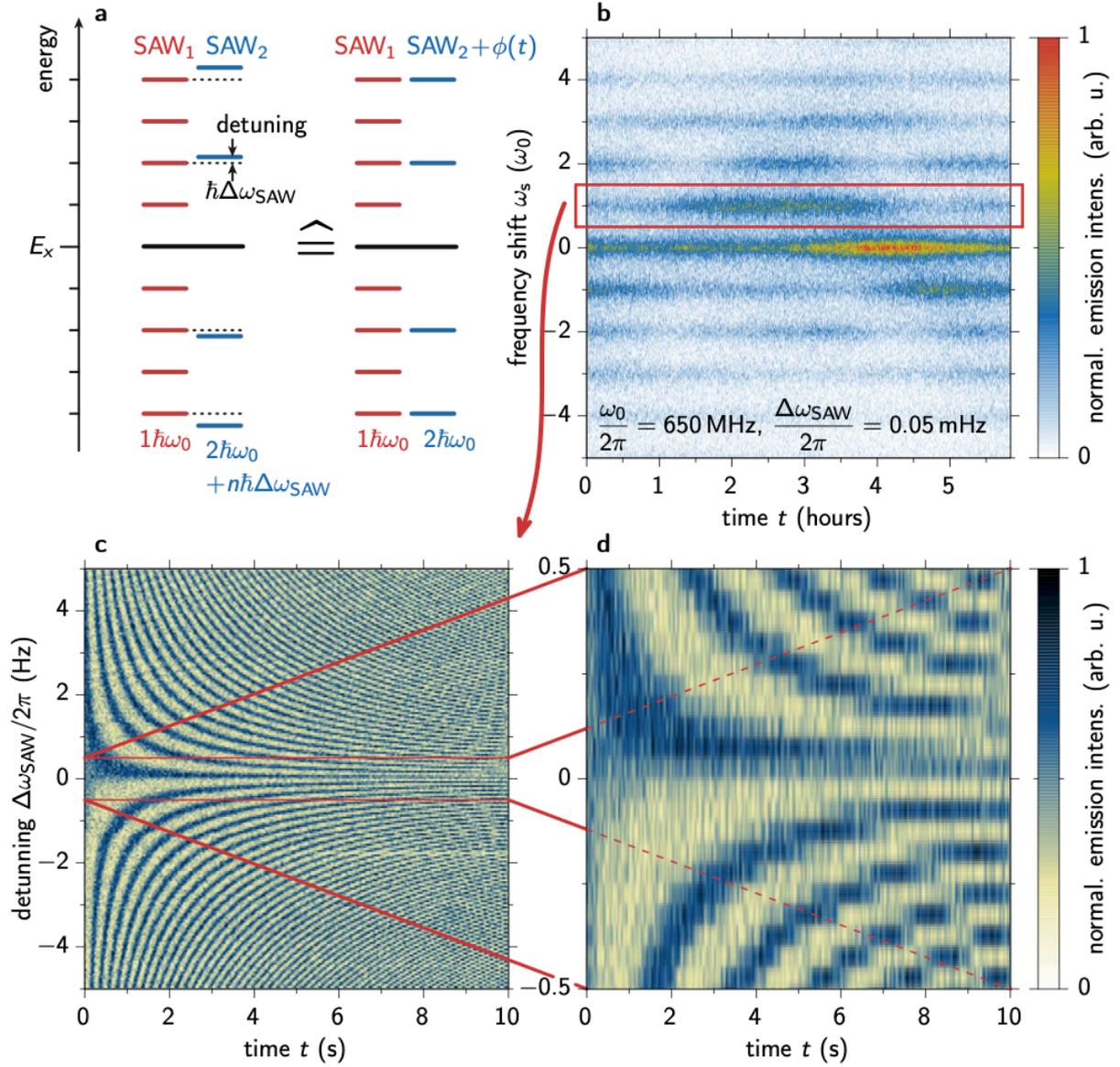

**Figure 4 | Frequency ring generation a,** Energy level diagram for finite detuning. The detuning $\Delta_{SAW}/2\pi$ gives rise to a time-dependent phase of the $\omega_{SAW} - 2\omega_{SAW}$ resonance condition. **b,** Time evolution of the PSB spectrum for $\Delta_{SAW}/2\pi = 50$ µHz. The observed period of the intensity oscillations matches the expected $T_\phi = 2\pi/\Delta_{SAW} \cong 5.55$ hours. **c,** Colour-coded intensity of the $+1 \cdot \omega_0$ PSB as a function of time and $\Delta_{SAW}/2\pi$ confirming the anticipated increase of the oscillation period as $|\Delta_{SAW}/2\pi|$ increases. **d,** Zoom-in to the central part of the data shown in **c** (red box) highlighting the central symmetry of the observed pattern confirming the faithful mapping of $\Delta_{SAW}/2\pi$ onto $T_\phi$.

Finally, we study the sensitivity of our hybrid nonlinear optomechanical coupling scheme to variations of mechanical frequency. To this end we deliberately introduce a small, yet finite detuning $\Delta\omega_{SAW}$ to the SAW such that $\omega_{SAW}^{(2)} = 2\omega_0 + \Delta\omega_{SAW}$. The resulting PSB alignment is shown schematically in Figure 4a. As depicted, introducing $\Delta_{SAW}$ is equivalent to a time-varying relative phase $\phi(t)$ of the $\omega_{SAW}^{(2)} = 2\omega_{SAW}^{(1)}$ nonlinear mixing scheme. Owning the precision of state-of-the-art rf electronics $\omega_{SAW}^{(1)}/2\pi = 0.65\,\text{GHz}$ and $\omega_{SAW}^{(2)}/2\pi = 1.30\,\text{GHz}$ can be set with a µHz resolution. Thus, the accessible sub-Hertz frequency range of $\Delta_{SAW}$ can easily be significantly smaller than the linewidth of the laser ($\delta\omega_{laser}/2\pi \leq 300\,\text{kHz}$) and that of the exciton transition ($\delta\omega_X/2\pi \approx 1\,\text{GHz}$). As proof-of-principle, we set $\Delta\omega_{SAW}/2\pi = 50\,\text{µHz}$, corresponding to a time-dependent phase $\phi(t) = \Delta\omega_{SAW}t$, and detect the scattered photon spectrum as a function of time over one full cycle of this oscillation. In Figure 4b, we plot the recorded intensity in false-colour representation as a function of time $t$ (horizontal axis) and the relative optical frequency shift $\omega_s$ (vertical axis). The experimental data exhibit a clear oscillation of the intensities of the PSB and the ZPL with a period of precisely $T_\phi = 2\pi/\Delta\omega_{SAW} \cong 5.55\,\text{hours}$. Remarkably, this demonstrated frequency rung is only $7.6 \cdot 10^{-14} \omega_{SAW}^{(1)}$ and $1.6 \cdot 10^{-19} E_X/\hbar$, underlining the outstanding stability of our hybrid non-linear optomechanical scheme. Again, the frequency rung of the sideband intensity is fully tunable via $\Delta_{SAW}$ which is studied in detail in Figure 4c. Here the intensity of the $+1 \cdot \omega_0$ PSB is plotted in false-colour representation as a function of time (horizontal axis) and the SAW detuning $\Delta_{SAW}/2\pi$ (vertical axis). As the detuning is varied between $-5\,\text{Hz}$ and $+5\,\text{Hz}$ the period of the oscillation detected in the time-domain faithfully changes accordingly, creating the pattern in Figure 4c that is inversion symmetric with respect to the $\Delta\omega_{SAW} = 0$ line. This pattern can be seen even more clearly in Figure 4d, which is a zoom-in on the regime of small detunings $-0.5\,\text{Hz} \leq \Delta\omega_{SAW}/2\pi \leq +0.5\,\text{Hz}$ of the data shown in Figure 4c. An instructive theoretical reproduction of this pattern is given in the Supplementary Information.

In conclusion, we have demonstrated optomechanical nonlinear three-wave mixing of two mutually coherent SAW fields and the optical field of a laser via the optical transition of a single QD. The phase matching underlying the interference enables deterministic enhancement or suppression of the different PSBs, which is faithfully detected in the resonantly scattered photon spectrum and accounted for by our theory. While our particular implementation is based on epitaxial QDs made of III-V compound semiconductors, many other systems like defect centres in diamond[19,21], silicon carbide[21] or two-dimensional materials[41] have already been proven to be well-suited to be interfaced with SAWs. Moreover, the driving optomechanical interaction can be drastically enhanced by embedding the quantum emitter in phononic or optomechanical resonators building on demonstrated monolithic integration on GaAs[10,42]. Ultimately, coherent microwave-optical quantum state transduction comes into reach. For example, a single SAW phonon emitted by a microwave superconducting qubit can be transmitted on-chip[18] to a single QD or another type of quantum emitter which converts it to a single photon employing the scheme demonstrated here. Furthermore, our SAW-modulation paves the way towards dynamically reconfigurable control of integrated quantum photonic devices[43–46], for instance for sideband-resolved single photon routing. Alternatively, the heterointegration of the semiconductor on $LiNbO_3$ harnesses the strong piezomechanical and nonlinear optical properties of the host substrate and the large optomechanical coupling of the compound semiconductor[47–49].

## Methods

### Sample design

The sample was grown by solid-source molecular beam epitaxy. It contains a single layer of In(Ga)As QDs embedded in a planar optical cavity formed by two distributed Bragg reflectors (DBRs) of 8 and 15 alternating layers of AlAs and GaAs on the top and bottom, respectively. The QDs were grown with low surface density to enable selective optical excitation of and detection from a single dot. The cavity resonance was designed to match the emission band of the QDs. The cavity quality factor of $Q \approx 150$ is sufficiently high to increase the light matter interaction. At the same time the corresponding cavity linewidth of $\Delta f \geq 2.2 \text{ THz}$ is sufficiently large to avoid dynamic modulation of the cavity resonance with an amplitude exceeding the cavity linewidth[50,51]. This ensures that we can exclude any time-modulation of the Purcell-effect as reported in our previous work on high-$Q$ cavities[52]. Multiharmonic IDTs (Ti 5 nm, Al 50 nm) were patterned by standard electron beam lithography in a lift-off process directly on top of the sample to facilitate SAW excitation at the fundamental frequency and three overtones. The IDTs were either fabricated with a constant wavelength of $\lambda = 9.91 \mu m$,[36] or, in order to realize frequency chirped transducers[53], of wavelength linearly varying from $\lambda_0 = 8.94 \text{ μm}$ to $\lambda_1 = 10.38 \text{ μm}$ along the length ($L_{\text{IDT}} = 400$μm) of the IDT. The first design allows SAW generation at a fundamental frequency of about 338 MHz and at overtones of 677 MHz, 1015 MHz and 1355 MHz, while the second design enables SAW generation across four frequency bands, nominally spanning from $300 - 350\,\text{MHz}$ (fundamental), $600 - 700\,\text{MHz}$ (first overtone), $900 - 1050\,\text{MHz}$ (second overtone) and $1200 - 1400\,\text{MHz}$ (third overtone). Additional details on the IDT design and characterization is provided in the Supplementary Material.

*Acousto-optical spectroscopy*

For measurement, the sample was mounted in a closed-cycle optical cryostat and kept at a temperature of $T = 5\,\text{K}$. Single QDs are optically accessed perpendicular to the sample surface/planar cavity by a confocal microscope setup and resonantly excited by a continuous wave frequency-tunable laser (linewidth $\delta\omega_{laser}/2\pi \leq 100\,\text{kHz}$). The resonance fluorescence signal is collected in the same direction and, in order to suppress reflected laser light, excitation and detection are cross-polarised with respect to each other. The second order correlation function $G^{(2)}(\tau)$ was recorded with a fiber-based Hanbury Brown & Twiss setup equipped with Si avalanche single photon detectors with a time resolution $< 300\,\text{ps}$. To resolve the PSB, the collected light is spectrally filtered using a piezo-tunable Fabry Perot etalon (free spectral range $FSR = 60\,\text{GHz}$, finesse $\mathcal{F} = 263$) and detected by a single photon avalanche photodetector (SPAD). The temperature of the etalon was stabilized using a proportional-integral-derivative (PID) control loop.

The electrical signals used to excite SAWs were provided by two independent signal generators, that allowed for tuning of signal amplitude, frequency and relative phase. To ensure a stable phase-lock between both SAWs, the two signal generators were referenced to a common $10\,\text{MHz}$ oscillator. Both output signals were added using a standard rf-power combiner and applied to the IDT[36]. For the data shown in the Supplementary Information, two IDTs of the same design were used to generate counterpropagating SAWs.

*Theory*

We describe the QD as a two-level system and obtain the resonance fluorescence by calculating the correlation function $G^{(1)}(t, t+\tau) = \langle \sigma_+(t)\sigma_-(t+\tau)\rangle$.[54] The QD is coupled to a classical light field using rotating wave and dipole approximations and the influence of the strain of the SAW waves is introduced as an energy-modulation of the transition energy $E_X$ via deformation potential coupling[55]. In addition, we consider the spontaneous recombination rate $\gamma$ of the exciton state. The temporal evolution is calculated using the Master equation in Lindblad form and the two-point correlation function $G$ is found by quantum regression. In the experiment, the resonance fluorescence signal emitted by the QD is spectrally filtered by an etalon. Using a Lorentzian filter function $F(\omega)$ (spectral width $\delta_e/2\pi = 0.41$ GHz for all simulations) we obtain for the scattered light intensity

$$I(\omega) = 2\text{Re} \frac{1}{T}\int_0^T \int_{-\infty}^{\infty} \int_0^{\infty} F^*(t-\tau)F(t-\tau-s)G(\tau,\tau+s)e^{i\omega s} \, ds \, d\tau \, dt,$$

where $F(t) = \exp(-\delta_e|t|)$ is the Fourier transform of $F$ and $T$ is the period of the SAW field. Further details are included in the Supplementary Information.

## Acknowledgements

This project has received funding from Deutsche Forschungsgemeinschaft (DFG, German Research Foundation) via the Emmy Noether Program (KR3790/2) and the Cluster of Excellence "Nanosystems Initiative Munich" (NIM). D.W. thanks the Polish National Agency for Academic Exchange (NAWA) for financial support within the ULAM program (No. PPN/ULM/2019/1/00064). K.M and J.J.F. acknowledge support by Deutsche Forschungsgemeinschaft (DFG, German Research Foundation) under Germany's Excellence Strategy – EXC-2111 – 390814868. K. M. thanks the German Federal Ministry of Education and Research (BMBF) via the funding program Photonics Research Germany (contract number 13N14846) and the Bavarian Academy of Sciences and Humanities (BAdW) for financial support. D.W., K.M., J.J.F., T.K. and P.M. acknowledge support from NAWA under an APM grant. M.W., M.N. and H.J.K. thank Achim Wixforth for his continuous support and invaluable discussions.

## Author contributions

H.J.K., D.W. and P.M. initiated the project. M.W. designed and fabricated devices, built experimental setup and performed experiments. M.W., D.W., M.N. and H.J.K. analysed the experimental data. D.W., P.M. and T.K. developed the theoretical model. K.M. advised on the design of the experimental setup. K.M. and J.J.F. performed crystal growth and pre-characterization of the semiconductor heterostructure. H.J.K., D.W., P.M. and M.W. wrote the manuscript with input from all authors.

## Additional information

Supplementary information is available in the online version of the paper. Reprints and permissions information is available online.

Correspondence be addressed to H.J.K.

## Data availability

The data used to generate the plots within this paper are available from the corresponding author upon reasonable.

## Competing financial interests

The authors declare no competing financial interests.